\documentclass[sigconf,screen]{myclass}
\usepackage[frozencache,cachedir=.]{minted}
\usepackage{dcolumn}
\usepackage[flushleft]{threeparttable}

\graphicspath{ {./case_study_plots/} }

\begin{document}
\title{An Improved and Extended Bayesian Synthetic Control}

\author{Sean Pinkney}
\affiliation{%
  \institution{Comscore, Inc.}
  \city{Reston}
  \country{Virginia}
}
\email{spinkney@comscore.com}

\begin{abstract}
An improved and extended Bayesian synthetic control model is presented expanding upon the latent factor model in \cite{tuomaala2019bayesian}. The changes we make include 1) standardization of the data prior to model fit - which improves efficiency and generalization across different data sets; 2) adding time varying covariates; 3) adding the ability to have multiple treated units; 4) fitting the latent factors within the Bayesian model; and, 5) a sparsity inducing prior to automatically tune the number of latent factors. We demonstrate the similarity of estimates to two traditional synthetic control studies in \cite{abadie2010synthetic} and \cite{abadie2015} and extend to multiple target series with a new example of estimating digital website visitation from changes in data collection due to digital privacy laws. 
\end{abstract}

\maketitle

\section{Introduction}
\label{sec:intro}
The paper by \cite{tuomaala2019bayesian} uses a Bayesian factor model to estimate the counterfactual treatment effect. The latent factors capture the pre-treatment information by pooling information across all the units - in contrast to the traditional synthetic control where the treatment effect is estimated by an optimal weighting of the non-treated units. The counterfactual treatment effect is estimated by including a time-varying parameter on a treatment indicator variable. The indicator is set to one for the treated unit(s) during the treatment period and zero for all other units and times. Setting this parameter to zero in the post-hoc analysis generates the counterfactual synthetic control. We find improvements in estimation speed, efficiency, and robustness by removing that extra treatment effect parameter and masking the treatment period data for the treated unit. This more closely aligns with the traditional synthetic control model, where the treated unit's data is only used for pre-period model estimation. 

Estimating the counterfactual outcome may be cast as a missing data problem \cite{athey2021matrix}. Although the untreated state is impossible to observe, we assume that the data generating process of the treated and untreated units is similar, which allows the estimation of the hypothetical untreated outcome on the treated. The Bayesian framework captures the joint posterior distribution across the data and parameters to estimate the missing data during the model fit\footnote{The missing data are treated as parameters to be estimated by HMC so they are estimated jointly with the data and parameters rather than drawn from a distribution.}. The latent factors capture the cross unit heterogeneity, seasonality, and other, potentially unobserved, effects. We avoid having to search for the optimal number of latent factors by using a sparsity inducing horseshoe prior to regularize the unimportant factors. This not only avoids an expensive optimal search but protects against overfitting. 

We show the similarity of this model to traditional synthetic control estimates in the West Germany reunification and the California tobacco case studies which are from \cite{abadie2010synthetic} and \cite{abadie2015} respectively. Lastly, we include a real-world business use case from Comscore, Inc. which occurred in the summer of 2020 on website collection changes for compliance with European privacy laws. The entire estimation is fit in the Stan modeling language \cite{stan}. 

We call attention to the simple step of pre-processing the data, which is, perhaps, an additional key improvement. Although it is a common precursor to model estimation, the code for the Bayesian latent factor model in \cite{tuomaala2019bayesian} does not perform this step. We find dramatic improvement in estimation and that this allows the model to generalize to different data sets with minimal adjustment. No treatment period information is used when scaling the outcome variable which protects against biasing the estimate of the treatment effect. For example, in the West Germany and California tobacco studies we shift the outcome data using the last data point just before the intervention period and then scale by the pre-intervention standard deviation.

\subsection{Bayesian Latent Factor Model}
Latent factor models for synthetic control predate \cite{tuomaala2019bayesian} and the most comprehensive study is in the interactive fixed effect (IFX) model in \cite{xu_2017}. The IFX model is a two-step estimation procedure that estimates the interactive fixed-effects for the control group in the first step. The second step captures the treated unit latent factors by estimating the their factor loadings in the pre-treatment period. A drawback of that approach is that the latent factors are fit only on control group data (for all time periods) and then the loadings are fit in a separate procedure, in only pre-treatment periods, potentially decreasing the efficiency of the estimates. Our proposed model uses more of the data for estimation - only withholding the data of the treated unit in the treated period - while simultaneously estimating the latent factors. To get variance estimates for the IFX estimate a parametric bootstrap procedure is used, whereas the Bayesian model outputs full uncertainty distributions for every parameter. 
 
We mask the treated unit(s) outcome in the intervention period and estimate it as a parameter to be fit in the model. All other data are included which allows the Bayesian estimation to impute the value from the estimated posterior distribution. As with traditional synthetic control, it is necessary that the "donor" series (see \cite{abadie2020}) be suitable candidates to the target series for plausible estimates either by domain knowledge or an additional framework such as in \cite{microsynth}. The model is composed of three core components. A latent factor component $F$ that includes post-multiplication by a random-vector of series traits $\beta$, a time specific component $\Delta$, and a series specific component $\kappa$. A covariate matrix $X$ is optionally supplied. The outcome matrix is $J \times T$ with $J$ series and $T$ time periods.

Each of the $J$ series is a row vector of $T$ times that follows
$$
y_j = F_j\beta_j + X_j\gamma + \Delta + \kappa_j, 
$$
where $F$ is a $T \times L$ matrix, $\beta$ is a $L$ length vector, $X$ is a matrix of covariates with $\gamma$ coefficients,  $\Delta$ is a size $T$ vector that is the same across all $J$ series, and $\kappa$ is a scalar for the $j^{th}$ series effect.

The Bayesian latent factor loadings and weight estimation procedure is specified in detail at \cite{farouni2015}. We make a simplifying assumption that the loading matrix $F$ is uncorrelated. The two-fold effect of this assumption is that computation speed increases as the $J$ series are fit with univariate normals and, more importantly, it obviates the need to estimate an additional correlation matrix, easing posterior geometry exploration of HMC, as there is already an implicit sharing of information across the series in $F$, $\beta$, and $\Delta$. To ensure identification we utilize the restrictions in \cite{farouni2015} where the upper triangular elements of $F$ are set to zero and the diagonal elements positive.

The priors for $F$, $\gamma$, $\Delta$, $\kappa$, and $\sigma$ are

$$
\begin{aligned}
F &\sim \mathcal{N}(1, \, 0) \\
\gamma &\sim \mathcal{N}(1, \, 0) \\
\Delta &\sim \mathcal{N}(0, \, 2) \\
\kappa &\sim \mathcal{N}(0, \, 1) \\
\sigma &\sim \mathcal{N}(0, \, 1).
\end{aligned}
$$
The prior for $\Delta$ is given a slightly wider normal as there are potentially larger deviations across time. All the series are given the same standard deviation $\sigma$ (i.e. a scalar) across time. This could easily be modified to use a $T$ length vector that varies across each unit of time or even a $J \times T$ matrix that varies across series and time. However, we find that with a vector $\sigma_t$ the model struggles with increased HMC rejections, tree-depth warnings, and  slow computation, indicating non-identifiability. Nevertheless, the assumption of homoskedastic errors across time may be unsuitable for other series, and the user may replace the above with a normal that varies across time or multivariate normal distribution that incorporates additional correlation across entities.

An additional horseshoe+ prior given to $\beta$ to regularize unnecessary factors. The horseshoe+ prior is shown to reduce lower mean squared error and achieves the optimal Bayes risk up to a constant \cite{bhadra2015horseshoe}. $\beta$ varies across the number of latent factors $l \in 1:L$ and the number of entities $j \in 1:J$,

$$
\begin{aligned}
(\beta_{l, \, j} \mid \lambda_l, \, \eta_j, \, \tau) &\sim \mathcal{N}(0, \, \lambda_l) \\
(\lambda_l \mid \eta_j, \, \tau) &\sim \mathcal{C}^+(0, \, \tau \eta_j) \\
\eta_j &\sim \mathcal{C}^+(0, \, 1) \\
\tau &\sim \mathcal{C}^+(0, \, 1).
\end{aligned}
$$
Although $\beta$ varies across $i$ and $j$, the local shrinkage parameter $\lambda$ only varies across the $L$ latent factors. We let the mixing parameter $\eta$ weight each of the series as a pseudo-synthetic weight. A half-Cauchy is added to the global shrinkage parameter $\tau$.  The half-Cauchy priors are reparameterized by setting the raw horseshoe+ priors as variates constrained on $(0, \, 1)$ then transformed to a half-Cauchy. For example, the parameter $\alpha$ follows a half-Cauchy if $\alpha_{\text{raw}} \in (0, \, 1)$ and $\alpha = \tan(\pi \alpha_{\text{raw}} / 2)$ \cite{betancourtcauchy}.  

The final outcome model is
$$
y_j \sim \mathcal{N}(F_j\beta_j + X_j\gamma + \Delta + \kappa_j, \, \sigma),
$$
where each of the $J$ series has $T$ time points that are distributed as normal with mean given by the length $T$ factor model, length $T$ time specific effects, and the scalar country specific effect, and with standard deviation $\sigma$.

\section{Applications}
In this section, we present several case studies to illustrate the use of the model. The German reunification and California tobacco program cases are now classic examples of synthetic control, first appearing in \cite{abadie2015} and \citep{abadie2010synthetic} respectively. We compare the proposed model to the synthetic control of the original models. If the original model code is unavailable we replicate according to the description in the paper. In the West Germany study the replication code is available at \cite{germanrep}, the California tobacco control program replication code is unavailable from the original authors, however, we use the R package \texttt{tidysynth}\cite{tidysynth} which has replicated the results according to the paper specifications. 

The last example is a business case study using Comscore, Inc. data. The data are monthly digital visitation data to news websites from November 2019 through September 2020. A drop in digital collections occurred in the summer of 2020 due to implementation of changes in compliance with European privacy laws. The model estimates the counterfactual visitation as if the changes had not occurred.  

\subsection{German Reunification and California Tobacco Program}

The first two examples were studied in  \cite{abadie2015} and \citep{abadie2010synthetic}, replicated in \cite{tuomaala2019bayesian}, and have been used as the principal examples of synthetic control (see \cite{engelbrektson2020} and \cite{tidysynth}). 

\subsubsection{German Reunification}
The German reunification data is from \cite{germanrep}. The replicated method from \cite{abadie2015} creates the synthetic control by solving a quadratic program in the R package \texttt{Synth} \cite{rsynth}\cite{r}.  

The data consist of GDP measured in Purchasing Power Parity (PPP)-adjusted 2002 USD of 17 OECD member countries from  1960 - 2003. The pre-intervention period is from 1960 to 1990 - at which point German reunification occurred. The $X$ predictors are average GDP, average inflation rate, average trade openness, average industry share of value added from 1981 - 1990, average percentage of secondary school attained in the total population aged 25 and older from 1980 - 1985, and average investment rate from 1975 - 1980. 

\begin{figure}
\includegraphics{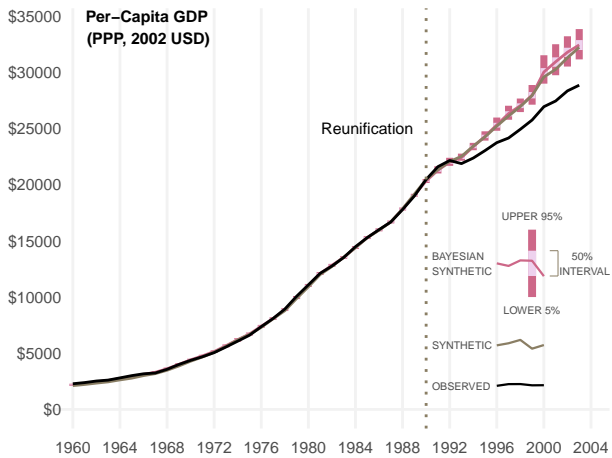}
\caption{Trends in per Capita GDP: West Germany, Bayesian Synthetic West Germany, and Traditional Synthetic West Germany}
\label{fig:germany_trend}
\end{figure}

We find minimal benefit to covariates that were added in the original model and leave them out. See the next two case studies for an examples of including covariates. We use 8 latent factors, double what \citep{tuomaala2019bayesian} use, to stress the sparsity inducing horseshoe+ prior. The model form is
$$
y_j \sim \mathcal{N}(F_j\beta_j + \Delta + \kappa_j, \, \sigma),
$$
where each $y_j$ is normalized by subtracting the 1989 value of $y_j$ and dividing by the 30 year pre-intervention standard deviation of $y_j$. The fit is performed using Stan with 4 parallel chains, using 500 warm-up iterations, and 500 post-warmup iterations. The initial value is set to 0.1 with \texttt{max\_treedepth} at $14$ and \texttt{adapt\_delta} set to 0.95. No divergences or tree-depth issues were noted. 

\begin{figure}
\includegraphics{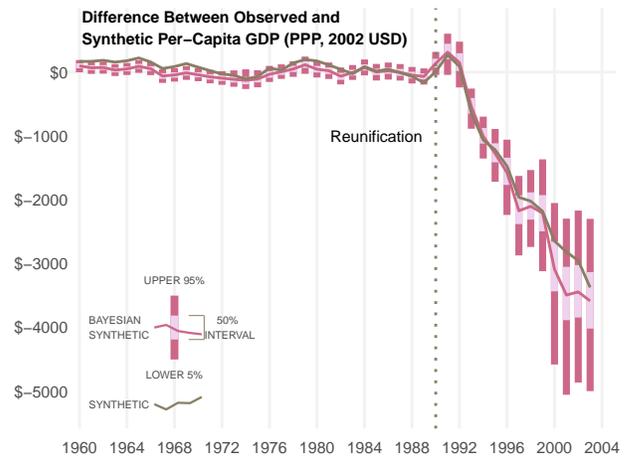}
\caption{Per Capita GDP Gap between West Germany, Bayesian Synthetic West Germany, and Traditional Synthetic West Germany}
\label{fig:germany_diff}
\end{figure}

The trend in GDP per capita of observed and model fits are in figure \ref{fig:germany_trend}. The two synthetic fits are similar. Both align closely with the observed GDP-per-capita values in the pre-intervention period. The post-intervention estimate is statistically indistinguishable up to year 2000 with the Bayesian model estimating a higher effect post-2000. The 90\% and 50\% probability intervals of the Bayesian estimate are shown. The true estimate is then believed to be in the interval with 90\% (50\%) probability. It is easier to see the difference of the two fits in figure \ref{fig:germany_diff}. This figure shows the difference of the synthetic trends in GDP-per-capita relative to observed trend, Bayesian probability intervals are included as well.

\subsubsection{California Tobacco Control Program}

The California tobacco control replication data is from the R package \texttt{tidysynth} \cite{tidysynth}. The replicated method is similarly from \texttt{tidysynth} and solves the quadratic program that specified in \cite{abadie2010synthetic}. The data are annual state-level panel data for 39 states from 1970 - 2000 of per-capita cigratte sales. The pre-intervention period is from 1970 to 1988 when proposition 99 was passed. The $X$ predictors are average retail price of cigarettes, logged per capita state personal income, the percentage of the population age 15 – 24, and per capita beer consumption averaged over the 1980 – 1988 period. Furthermore, three years of lagged smoking consumption (1975, 1980, and 1988) are included, which mimics exactly the inputs to the original model.

The model form is
$$
y_j \sim \mathcal{N}(F_j\beta_j + X_j\gamma + \Delta + \kappa_j, \, \sigma),
$$
where each $y_j$ is normalized by subtracting the 1987 value of the packs-per-capita value and dividing by the pre-intervention standard deviation of $y_j$ (a period of 28 years). The fit is performed using Stan with 4 parallel chains, 500 warm-up iterations, and 500 post-warmup iterations. The initial value is set to 0.1 with \texttt{max\_treedepth} at $13$ and \texttt{adapt\_delta} set to 0.8. No divergences or tree-depth issues were noted. 

The included covariates align the Bayesian latent factor model closer to the quadratic program fit. Each of the covariates are static variables, meaning they do not fluctuate across time. It is possible to include time-varying covariates and other adjustments in the proposed model (see the following case study), as opposed to the models in the \texttt{Synth} package. A frequentist synthetic control with time varying covariates is possible in such newer \texttt{R} packages such as \texttt{microsyth}\cite{microsynth} and \texttt{gsynth}\cite{xu_2017}. The only covariate with a standardized effect different from zero is the lagged smoking consumption of 1988. However, be cautious of overly interpreting these coefficients as the meaning is complicated by the standardization of both the outcome variable - cigarette sales in packs - and the predictors\footnote{The transformation back to the original units involves the ratio of standard deviations of $y$ and $x$ as in $\gamma^* = \gamma \frac{\texttt{sd}(y)}{\texttt{sd}(X)}$. In fact, transforming the effect of cigarette sales in 1988 into the original units for California gives an estimate of -0.5, or a decrease of 0.5 pack per capita for every increase in 1988 cigarette sales.}. Note that the standardization of each state outcome across time means that the effect of the predictors implicitly varies across states. 

\begin{figure}
\includegraphics{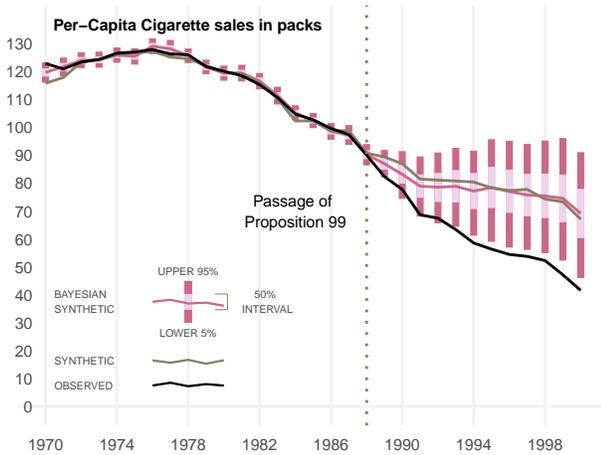}
\caption{Trends in per Capita Cigarette Sales: California, Bayesian Synthetic California, and Traditional Synthetic California}
\label{fig:smoking_trend}
\end{figure}

%\begin{figure}
%\includegraphics{smoking_cov.eps}
%\caption{Standardized Covariate Effects with 50\% probability interval (inner) and 90\% probability interval (outer). The line at 0 indicates no effect.}
%\label{fig:germany_diff}
%\end{figure}

\begin{table}[ht]
\begin{threeparttable}[b]
\caption{California Proposition 99: Standardized Predictor Coefficients}
% \centering
\begin{tabular}{rlrrrrr}
  \\[-4ex]\hline 
\hline \\[-1.8ex] 
 & Variables & Mean & SD & $\hat{R}$ & $\text{ESS}_\text{Bulk}$ & $\text{ESS}_{\text{Tail}}$ \\ 
  \hline
	1 & Log Income        & -0.05 & 0.21 & 1.02 & 707 & 966 \\ 
  	2 & Retail Price Cig. & -0.05 & 0.22 & 1.01 & 789 & 899 \\ 
  	3 & \% Age 15 to 24   & -0.02 & 0.18 & 1.00 & 1056 & 1032 \\ 
  	4 & Beer Sales        & 0.03 & 0.20 & 1.00 & 1088 & 1134 \\ 
  	5 & Cig. Sales 1975   & 0.49 & 0.50 & 1.00 & 1253 & 1423 \\ 
  	6 & Cig. Sales 1980   & 0.37 & 0.58 & 1.00 & 1415 & 1499 \\ 
  	7 & Cig. Sales 1988   & -1.02 & 0.37 & 1.01 & 751 & 978 \\ 
   \hline\\[-1.8ex]
\end{tabular} 
\vspace{-1ex}
\begin{tablenotes}
\item \scriptsize \textit{Note:} All predictors except \% Age 15 - 24 standardized from their original scale. 
The proportion of 15 - 24 year-old's is logit transformed prior to standardization.
\end{tablenotes}
\end{threeparttable}
\end{table}

\begin{figure}
\includegraphics{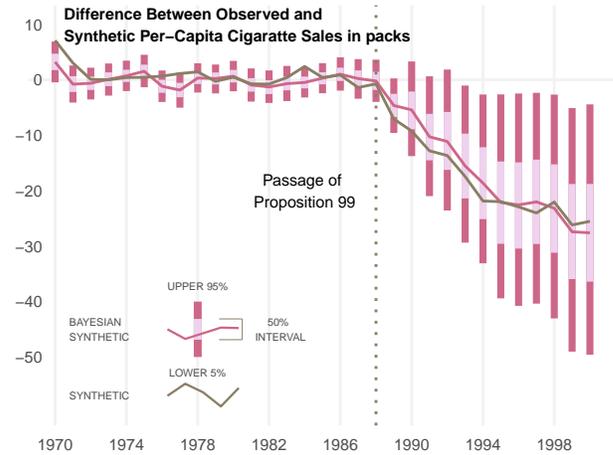}
\caption{Per Capita Cigarette Sale Gap between California, Bayesian Synthetic California, and Traditional Synthetic California}
\label{fig:smoking_diff}
\end{figure} 
 
Figure \ref{fig:smoking_trend} shows the trend of cigarette packs per capita on observed California and the two synthetic California fits. The mean of the two fits are similar with both following the pre-intervention variability of California closely. The post-intervention estimate is slightly lower for the Bayesian model. The Bayesian probability interval indicates a wide variability in the possible effect of the law. The difference is illustrated more clearly in figure \ref{fig:smoking_diff} which shows the difference from the observed data to the fits. The counterfactual effect increases with time as does the uncertainty - beginning with an uncertainty range of $0 - 10$ packs per capita and ending at $5 - 50$ packs per capita .  

\subsection{Digital Data Collections}

We study the impact of a set of changes in Comscore\footnote{Comscore is a trusted currency for planning, transacting, and evaluating media across platforms. See \url{https://www.comscore.com/}.} digital collections that resulted in the partial loss of data for the UK, Spain, and France in the summer of 2020. As required by the EU General Data Protection Regulation (GDPR)\cite{gdpr}, Comscore relies on user consent for measurement of digital media consumption (including page visitation, app usage, video viewing as well as other interactions with digital content and advertising) in all European Union countries as well as the United Kingdom. If consent is not provided, Comscore obfuscates the personal data (e.g. browser cookie or device ID) associated with the measurement event. The software changes were rolled out in phases beginning in June 2020 and completed by August 2020. Interestingly, the phase-in occurred in June for the UK and France but not until July in Spain. The June collections showed a steeper decline than anticipated and worsening through July and August. A software issue was responsible and a correction was implemented in September of 2020. 

\begin{figure}
\includegraphics[width=3.3in, height=3.5in]{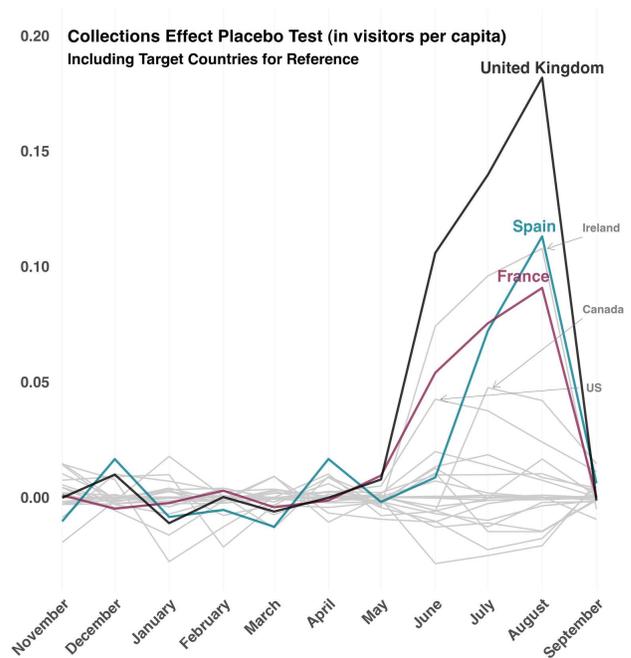}
\caption{Digital Collections Issue June 2020 - August 2020}
\label{fig:digital_placebo}
\end{figure}

In place monitoring easily caught the anomaly, however, the monitoring system relies upon human intervention to produce the counterfactual correction - that is, the estimate of what \textit{would} have been if the anomaly had \textit{not} occurred. The proposed model produces this additional piece of information  along with an uncertainty interval to arm analysts with better corrective tools. Furthermore, running a so-called "placebo" study - where each donor series is modeled as \textit{if} they were given the treatment to produce a placebo estimate - can aide in flagging anomalous series that were undetected from univariate time series monitoring mechanisms. If the observed data exhibit large deviations from the placebo estimate, this is a strong indication that the series is encountering anomalous issues. We show an example of this process on the digital case study in figure \ref{fig:digital_placebo}. 

\begin{figure}
\includegraphics[width=3.3in,height=3.75in]{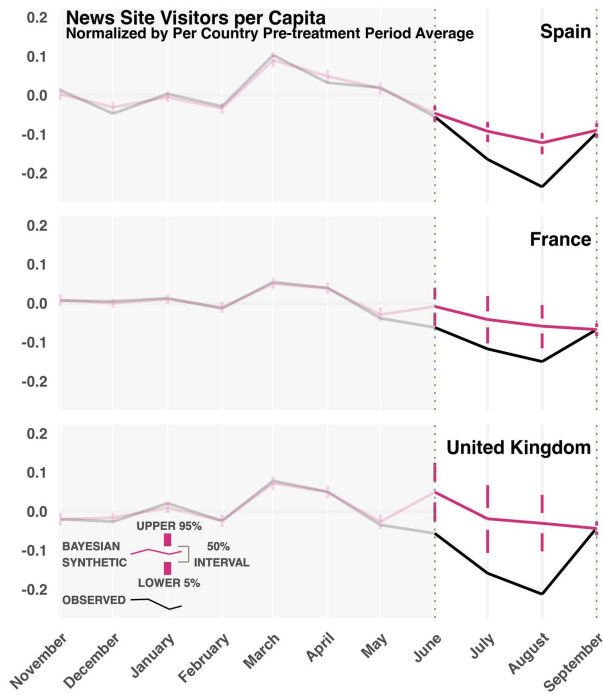}
\caption{News Site Visitors per Capita with a collections issue from June 2020 to August 2020}
\label{fig:digital}
\end{figure}

We illustrate the model on a subset of reported Comscore data - news and information sites from November 2019 through September 2021. In addition to the three affected countries, there are 21 donor countries\footnote{The donor countries are Argentina, Australia, Brazil, Canada, Chile, China, Colombia, Germany, India, Indonesia, Ireland, Italy, Japan, Malaysia, Mexico, New Zealand, Peru, Vietnam, United States, Sweden, and Switzerland.}. The outcome variable of interest is the number of unique visitors per capita. A time series covariate of the number of page views per capita is included. Since the collections issue did not effect the number of page views, the covariate is included for every country and time period. The country level population data is from \cite{popdata}. We use 10 latent factors, 500 warmup samples, 500 post-warmup iterations, 4 parallel chains, maximum tree depth of 13, and adapt delta of 0.99. Note that the treatment period is bounded between June and August and the final time point in September 2020 is observed for all countries.

Figure \ref{fig:digital} shows the visitor per capita series normalized by the pre-collections issue country average (8 month period) for the observed and the synthetic countries. The model shows a modest impact beginning in June with the largest impact in August, which is expected given the phased roll-out of the changes. The effects were most predominant in the UK. Furthermore, the uncertainty intervals give more insight what may be occurring to each affected country. For example, no changes were made to collections for Spain in June and the model estimates close to the observed value. The July and August effects in Spain were compounded by unrelated collection issues that additionally reduced the observed number of visitors. The uncertainty intervals are smaller, reflecting more certainty that the true visitation per capita is higher than observed.   

The placebo study is shown in \ref{fig:digital_placebo}. The model was fit on each of the donor countries to understand the sensitivity of the counterfactual estimate. The data for the collections period was withheld and the model fit as if the donor country was affected by the collections issue. Ireland shows an effect as great as Spain, which is not surprising, given how country traffic is assigned and the similarity and proximity of Ireland to the UK. Country traffic is based on a number of factors including IP addresses, which can bleed across country borders. The US and Canada show modest differences from observed, a possible indication of a larger, global effect or, perhaps, that English language sites were affected more severely.  
  
\section{Discussion}
This paper gives a number of improvements to the Bayesian synthetic control model that make the model more practical to implement. The model is shown to give similar results to the frequentist synthetic control model and a new case study on digital collections data is presented. Although data standardization is often assumed or  common practice, we find that it is necessary to achieve good model fits. It  improves the sampling efficiency in Stan while generalizing the model so that it may be used on different data sets without changing the priors.

A common pain point of latent factor models is choosing the number of factors and how to fit the factor model. In \cite{tuomaala2019bayesian} the latent factors are specified a priori and fit in PCA. We find that the additional computational complexity to run the latent factor model in Stan is minimal. In fact by including the latent factor fit in the model the uncertainty intervals include the extra variance from estimating these additional parameters. 

We specify a horseshoe+ prior that allows the practitioner to set a larger than anticipated number of latent factors and not worry about overfitting. This reduces either the ad-hoc nature of selecting the number of factors by having the model detect the relevant factors, or an expensive cross-validation procedure that would otherwise be needed to select the number of factors. We find that even by increasing the number of factors the horseshoe+ prior protects the model from overfitting and other fit issues. The main issue with increasing the factors is the increase in parameters further increases the time it takes to run the model.

Future research on the topic includes generalizations to the model such as hierarchical pooling, multivariate time series, or both. Estimation of the series covariance matrix is a potentially valuable addition, as it would quantify the cross series connection offering additional insight into understanding the counterfactual effect. Autoregressive effects may be added by lagging the latent factors. These effects can be used to study the decay of certain time varying effects, such as in advertising adstock models.    

\begin{acks}
The author would like to thank Gaurav Shahane and Tyler Morrison for their helpful discussions on the topic and Anjali Gupta for providing the context and data on the digital data case study. 
\end{acks}

%\section{Appendix: Number of latent factors is known}
%The parameters $\beta_{\mu}$ and $\beta_{\sigma}$ are $L$ length hyperprior vectors of the mean and standard deviation, respectively, of the normal distribution that characterizes $\beta$. However, we use a non-centered parameterization for $\beta$ and the $J \times L$ matrix $\beta^{*}$ is the standard normal offset. The final $\beta$ is derived from the above by $\beta_j = \beta_{\mu} + \beta_j^* \odot \beta_{\sigma}$ where $\odot$ means element-wise multiplication. 

\section{Appendix: Example Stan Code}
Replication Stan code for the German reunification study. 
\begin{minted}[mathescape,
               numbersep=5pt,
               gobble=2,
               frame=lines,
               framesep=2mm]{stan}
  functions {
   matrix make_F (int T, 
                  vector diagonal_loadings, 
                  vector lower_tri_loadings) {
    
    int L = num_elements(diagonal_loadings);
    int M = num_elements(lower_tri_loadings);
    matrix[T, L] F;
  
    int idx = 0; // Index for the lower diagonal 
    
    for (j in 1:L) {
      F[j, j] = diagonal_loadings[j];
      for(i in (j + 1):T) {
        idx += 1;
        F[i, j] = lower_tri_loadings[idx];
      }
    }
    
    for (j in 1:(L - 1)) {
      for (i in (j + 1):L) F[j, i] = 0;
    }
    
    return F;
  }
    matrix make_beta (int J, matrix off, 
                      vector lambda, 
                      real eta, 
                      vector tau) {
      
      int L = cols(off);
      vector[L] cache = ( tan(0.5 * pi() * lambda) * 
                          tan(0.5 * pi() * eta) );
      vector[J] tau_ = tan(0.5 * pi() * tau);
      matrix[J, L] out;
 
     for (j in 1:J)
        out[j] = off[j] * tau_[j];
        
    return diag_pre_multiply(cache, out'); 
   }
  }
  data {
   int T;              // times
   int J;              // countries
   int L;              // number of factors
   int P;
   matrix[P, J] X;     // predictors
   row_vector[T] Y[J]; // data matrix of order [J,T]
   int trt_times;
  }
  transformed data {
   int<lower=1> M = L * (T - L) + L * (L - 1) / 2;
   row_vector[J] j_ones = rep_row_vector(1, J);
   vector[T] t_ones = rep_vector(1.0, T);
  
   matrix[J, P] X_std;

   vector[J] y_mu;
   vector[J] y_sd;
   row_vector[T] Y_scaled[J];
   row_vector[T - trt_times] Y_pre_target;
  
   vector[P] x_mu;
   vector[P] x_sd;
  
   for (j in 1:J) {
    y_mu[j] = Y[j, T - trt_times];
    y_sd[j] = sd(Y[j, 1:T - trt_times]);
    
    Y_scaled[j] = ( Y[j] - y_mu[j] ) / y_sd[j];
  }
 
   for (p in 1:P) {
    x_mu[p] = mean(X[p]);
    x_sd[p] = sd(X[p]);
    X_std[, p] = ( X[p]' - mean(X[p]) ) / sd(X[p]);
  }
  
   Y_pre_target = Y_scaled[1, 1:T - trt_times];
  }
  parameters{
   vector[P] chi;       //  non-time varying predictors
  
   vector[T] delta;     // year fixed effects
   row_vector[J] kappa; // country fixed effects
  
   matrix[J, L] beta_off;
  
   vector<lower=0, upper=1>[L] lambda;
   real<lower=0, upper=1> eta;
   vector<lower=0, upper=1>[J] tau;

   row_vector[trt_times] Y_post_target;
   real<lower=0> sigma;
  
   vector<lower=0>[L] F_diag;
   vector[M] F_lower;
  }
  transformed parameters {
   matrix[L, J] beta = make_beta(J, 
                                 beta_off, 
                                 lambda, 
                                 eta, 
                                 tau);
  }
  model {
   chi ~ std_normal();
   to_vector(beta_off) ~ std_normal();
 
   F_diag ~ std_normal();
   F_lower ~ normal(0, 2);
  
   delta ~ normal(0, 2);
   kappa ~ std_normal();
   sigma ~ std_normal();

  {
    vector[J] predictors = X_std * chi; 
    matrix[T, L] F = make_F(T, F_diag, F_lower);

    row_vector[T] Y_target[1];
    row_vector[T] Y_temp[J]; 
    
    Y_target[1] = append_col(Y_pre_target,
                             Y_post_target);
    Y_temp = append_array(Y_target, 
                          Y_scaled[2:J]);
    
    for (j in 1:J)
      Y_temp[j]' ~ normal_id_glm(F, 
                                 delta + kappa[j] + 
                                 predictors[j], 
                                 beta[ , j], 
                                 sigma);
   }
  }
  generated quantities {
   vector[T] synth_out[J];
   matrix[T, L] F_ = make_F(T, F_diag, F_lower);
   matrix[T, J] Synth_ = F_ * beta +  
                         delta * j_ones  +  
                         t_ones * 
                         (kappa + (X_std * chi)');
   
   for (j in 1:J) 
     synth_out[j] = Synth_[, j] *  y_sd[j] + y_mu[j]; 
  }
\end{minted}
\newpage

\bibliographystyle{ACM-Reference-Format}
\bibliography{factor_model}

\end{document}